\documentclass[prd,amsmath,notitlepage,twocolumn]{revtex4-1}
\usepackage{graphicx}
\usepackage{float}
\usepackage{epstopdf,cancel}
\usepackage{epsf,latexsym,bbm,euscript}
\usepackage{amssymb,amsmath}
\usepackage{mathtools} 
\usepackage{times,graphics}
\usepackage{soul,xcolor}
\usepackage{mathtools}

\def\6{{\langle}}
\def\9{{\rangle}}
\newcommand{\defeq}{\vcentcolon=}

\newcommand{\be}{\begin{equation}}
\newcommand{\ee}{\end{equation}}
\newcommand{\ba}{\begin{eqnarray}}
\newcommand{\ea}{\end{eqnarray}}

 \newcommand{\mA}{{\mathrm{A}}}

\def\half{{\tfrac{1}{2}}}

\def\pad{{\partial}}

\def\sg{\textsl{g}}

\def\cO{\mathcal{O}}

\usepackage{url,hyperref}
\hypersetup{colorlinks,linkcolor={blue!55!black},citecolor={red!50!black},urlcolor={blue!45!black}}

\begin{document}

\title{The lid of Pandora's box: geometry near the apparent horizon  }

\author{Daniel R. Terno}
\affiliation{Department of Physics \& Astronomy,
 Macquarie University, Sydney NSW 2109, Australia}
   \affiliation{Shenzhen Institute for Quantum Science and Engineering,  Department of Physics, Southern University of Science and Technology, Shenzhen 518055, Guandong,  China}

\begin{abstract}
  Trapped regions bounded by horizons are the defining features of black holes.  However, formation of a singularity-free apparent horizon in finite time of a distant observer is consistent only with
  special states  of  geometry and matter in its vicinity. In spherical symmetry such horizons exist only in two classes of
  solutions of the Einstein equations. Both violate the null energy condition (NEC) and allow for expanding and contracting trapped regions. However, an expanding trapped region leads to a  firewall.
   The weighted time average of the  energy density  for an observer crossing this firewall
   is negative and exceeds the maximal  NEC violation that quantum fields can produce. As a result, either black holes only can evaporate or the semiclassical physics breaks down already at the horizons.
   Geometry of a contracting trapped region
  approaches   the ingoing Vaidya metric with decreasing mass. 
   Only one class of solutions allows  for a test particle to cross the apparent horizon, and for  a thin shell to collapse into a black hole. These results  significantly constrain  the   regular black hole models.
   Models with regular matter properties at the horizon can be realized only if significant departures from the semiclassical physics occur already at the horizon scale.
   The Hayward-Frolov   model may describe only evaporation, but not formation of a regular black hole.
 \end{abstract}
\maketitle

\section{Introduction}

Black holes are probably the most celebrated prediction of   classical general relativity (GR) \cite{he:book,fn:book}. Absence of a fully-developed theory of quantum gravity leaves  us with
 a hierarchy of approximate models  that combine GR and quantum mechanics  \cite{hv:book}.   
Quantum effects are known to modify the Einstein equations \cite{hv:book,bd:82,geft,rv:book} and to enable violations of the
classical energy conditions \cite{he:book,fn:book,hv:book,mmv:17,ks:20}. These results make plausible  three distinct  types  of ultra-compact objects (UCOs)   models
that purport to describe the observed astrophysical black holes \cite{cp:rev,bh-map}.

Models of the first type have   event horizon and singularity, even if their formation and properties are modified by quantum effects.
The final result of their evolution may be a black hole remnant \cite{coy:15}. Another scenario envisages formation of horizonless UCOs.
  The third option is a   black hole that has an apparent horizon, but  no event horizon or singularity.

Current observations  \cite{bambi:b,gw-12:rev,eht-1:19}  only weakly constrain these scenarios.  Given the apparent tension between quantum mechanics and GR, issues
 of logical consistency of models, as well as the information loss problem \cite{rev-0,rev-1}, it is important to understand what each scenario entails.

There are different opinions on  what makes a UCO a black hole \cite{curiel}. However, the strongest degree of consensus is that it should have a trapped {spacetime} region, whose boundary is
  the  apparent horizon \cite{he:book,fn:book,bambi:b}.  A trapped region is a domain where both ingoing and outgoing  future-directed null
  geodesics emanating from a spacelike  two-dimensional surface  with spherical topology have negative expansion \cite{he:book,fn:book,faraoni:b}.  The apparent horizon is the outer
boundary of the trapped region and the defining feature of a physical black hole (PBH)  \cite{frolov-do,bmmt:18}.  To be physically relevant the apparent horizon should form in a finite time of a distant observer.

Here we investigate the consequences of having  a PBH.
The simplest setting to investigate is a spherically-symmetric collapse, where the apparent horizon is unambiguously defined
in all foliations that respect this symmetry \cite{aphor}. 

 Building on the results of Refs.~\cite{bmmt:18,bmt:18,t:19}  we  describe the two possible classes of the   near-horizon geometries, discuss their properties, and consider the implications
  for singularity-free black hole models.

\section{Geometry near the apparent horizon: two classes of solutions}
 We    assume validity of semiclassical gravity. That means  we use classical notions (horizons, trajectories, etc.),
 and describe dynamics via the Einstein equations $G_{\mu\nu}=8\pi T_{\mu\nu}$, where the standard left-hand side is
  equated to the expectation value $T_{\mu\nu}=\6\hat{T}_{\mu\nu}\9_\omega$ of the renormalized energy-momentum tensor (EMT). The latter represents  both the collapsing matter and the created
excitations of the quantum fields. We do not assume existence of the Hawking radiation or specific properties of the state $\omega$.

Boundaries of the trapped region are nonsingular in   classical GR \cite{he:book,fn:book}, a requirement that is typically assumed to extend to the semiclassical regime.
 We implement this property  by  requiring that the scalars  $\mathrm{T}\defeq T^\mu_{~\mu}$ and $\mathfrak{T}\defeq T^{\mu\nu}T_{\mu\nu}$  are finite.  The Einstein equations
   imply that $64\pi^2\mathfrak{T}=R_{\mu\nu}R^{\mu\nu}$ and
   $8\pi\mathrm{T}=-R$, where $R_{\mu\nu}$ and $R$ are the Ricci tensor and the Ricci scalar, respectively.
    Finite values of these   scalars are a necessary regularity condition, and additional tests may be required.

    A general spherically-symmetric metric   in the  Schwarzschild coordinates  is given by
\be
ds^2=-e^{2h(t,r)}f(t,r)dt^2+f(t,r)^{-1}dr^2+r^2d\Omega, \label{sgenm}
\ee
where $r$ is the areal radius.  The Misner-Sharp mass  \cite{bambi:b,faraoni:b,ms} $C(t,r)$ is invariantly defined  via
\be
1-C/r\defeq \pad_\mu r\pad^\mu r  , \label{defMS}
\ee
and thus the function $f(t,r)=1-C(t,r)/r$ is invariant under general coordinate transformations. The apparent horizon is located at the Schwarzschild radius $r_\sg$ that is the largest root of
 $f(t,r)=0$ \cite{faraoni:b,aphor}.
The function $h(t,r)$   may contain   information about potential hairs of the stationary PBHs \cite{fn:book,cch:12}, and plays the role of an integrating factor in the coordinate transformations.

   It is convenient to introduce
\be
 \tau_t\defeq e^{-2h} T_{tt}, \qquad \tau^r\defeq T^{rr}, \qquad \tau^r_{t} \defeq e^{-h}T_t^{~r},
\ee
and represent the Misner-Sharp mass as
 \be
 C=r_\sg(t)+W(t,r-r_\sg), 
 \ee
  where the definition of the apparent horizon implies \be
  W(t,0)=0, \qquad W(t,x)<x.
  \ee
In this notation the three  Einstein equations for $G_{tt}$, $G_r^{\,t}$, and $G^{rr}$ become
 \begin{align}
&\frac{\pad_r C }{r^2}=8\pi \frac{\tau_t}{f}, \label{gtt}\\
&\frac{\pad_t C}{r^2}=8\pi e^h\tau_t^{r}, \label{gtr}\\
&\frac{\pad_r h}{r }=4\pi \frac{(\tau_t+\tau^r)}{f^2}, \label{grr}
\end{align}
 respectively.

   Regularity of the apparent horizon is expressed as a set of conditions on the potentially divergent  parts of the curvature scalars. For $\mathrm{T}$ and $\mathfrak{T}$ these are
\begin{align}
 \mathrm{T} &=(\tau^r-\tau_t)/f\to g_1(t) f^{\kappa_1},  \label{fin1}\\
 \mathfrak{T}&= \big((\tau^r)^2+(\tau_t)^2-2(\tau^r_{t})^2\big)/f^2  \to g_2(t) f^{\kappa_2},      \label{fin2}
\end{align}
for some $g_{1,2}(t)$ and $\kappa_{1,2}\geqslant 0$.   Here we exclude $T^\theta_{~\theta}\equiv T^\phi_{~\phi}$ from  consideration,
 because the Einstein equations imply  that
 $T^\theta_{~\theta}$  is finite (Appendix \ref{apreg}). 

 Eqs.~\eqref{fin1} and \eqref{fin2} require    the  EMT components to scale as some power $f^k$ as $r$ approaches the apparent horizon at $r_\sg$.
  All spherically-symmetric PBH solutions can be classified by the  values of $k$. Only the classes  $k=0$ (that results in the divergent energy density and pressure at the
                                                       apparent horizon) and $k=1$ (finite non-zero values of  energy density and pressure at the apparent horizon) are self-consistent.
 They are described below.

Using the advanced and the retarded null coordinates  allows additional insights into the near-horizon geometry. Its description in the terms of
 the advanced null   coordinate $v$,
 \be
dt=e^{-h}(e^{h_+}dv- f^{-1}dr), \label{intf}
\ee
is useful in the case of contracting apparent horizon, $r'_\sg<0$. A general spherically-symmetric metric in $(v,r)$ coordinates is
\be
  ds^2=-e^{2h_+}\left(1-\frac{C_+}{r}\right)dv^2+2e^{h_+}dvdr +r^2d\Omega. \label{lfv}
  \ee
If  $r'_\sg>0$  it is useful to employ the retarded null coordinate $u$.

Imposing the  finiteness conditions on the
Ricci scalar $R$ (Appendix \ref{rvr}) 
 at the apparent horizon $r_+(v)\equiv C_+(v,r_+)=r_\sg(t)$,
  we obtain that as $r\to r_\sg\equiv r_+$,
\begin{align}
    C_+(v,r)&=r_+(v)+w_+(v)(r-r_+)+w_2^+(r-r_+)^2\ldots, \\
 h_+(v,r)&=\chi_+(v)(r-r_+)+\ldots,
 \end{align}
for some functions $w_+$, $w_2$ and $\chi_+$, where the condition $w_+\leqslant1$ follows from the requirement $C<r$ outside the Schwarzschild radius.

Components of the EMT are related by \begin{align}
&\theta_v\defeq e^{-2h_+}\Theta_{vv}=\tau_t,  \label{thev}\\
&\theta_{vr}\defeq e^{-h_+}\Theta_{vr}=(\tau_t^r-\tau_t)/f, \label{thevr}\\
&\theta_r\defeq    \Theta_{rr}=(\tau^r+\tau_t-2\tau^r_t)/f^2,   \label{ther}
\end{align}
where $\Theta_{\mu\nu}$ denote the EMT components in $(v,r)$ coordinates.
The
 limits $\theta^+_{\mu\nu}\defeq\lim_{r\to r_+}\theta_{\mu\nu}$ are
\be
\theta_v^+= (1-w_+)\frac{r_+'}{8\pi r_+^2}, \qquad \theta_{vr}^+=- \frac{w_+}{8\pi r_+^2}, \qquad \theta_r^+=\frac{ \chi_+}{4\pi r_+}.        \label{the3}
\ee

\subsection{Divergent density and pressure}\label{divp}
Regularity conditions of Eqs.~\eqref{fin1} and \eqref{fin2} require  that divergent terms in the curvature scalars must cancel. Adding the requirement that
the function $C(t,r)$ is a real solution of Eq.~\eqref{gtt} results in
\be
 \tau_t=\tau^r=-\Upsilon^2(t)f^k, \qquad \tau^r_t=\pm \Upsilon^2 f^k,  \label{taus}
 \ee
 where $\Upsilon^2(t)$ is some function of time, $k<1$, and the higher-order terms are omitted. In the
 orthonormal basis the $(\hat t\hat r)$  block of the EMT  is
    \be
T_{\hat a \hat b}=-\Upsilon^2 f^{k-1} \!\!\begin{pmatrix}
1& \pm 1 \\
\pm 1 & 1 \end{pmatrix}.
\ee
 The upper (lower) signs of $T_{\hat t \hat r}$ correspond to growth (evaporation) of the PBH.     Leading terms of
the solutions   for $C(t,r)$ and $h(t,r)$     are given in Appendix \ref{solk}.      Static non-vacuum solutions with $\tau_t^r\to 0$
 are impossible for $k<1$, as the regularity condition Eq.~\eqref{fin2} cannot be satisfied unless all three components are zero.

    The null energy condition (NEC)   requires  $T_{\mu\nu} l^\mu l^\nu\geqslant0$ for all null vectors $l^\mu$. It is violated for all values
    of $k<1$ by radial vectors $l^{\hat a}=(1, \mp 1,0,0)$ for the evaporating and the accreting PBH solutions, respectively

 Violations of the NEC are bounded by quantum energy inequalities (QEIs) \cite{few:17,ks:20}.
  For a growing PBH, $r_\sg'>0$,
 in the reference frame of an infalling massive test particle the energy density
 (as well as the pressure and the flux), diverge (Appendix \ref{firewall}). Such a transient firewall leads to a violation  of the  quantum energy inequality \cite{ks:20,eleni},
  that is shown by repeating the analysis of Ref~\cite{t:19}. Henceforth we consider only the solutions with $r_\sg'<0$.

Comparison of Eqs.~\eqref{thev} -- \eqref{the3} with Eq.~\eqref{taus} shows that only the case $k=0$ is allowed, with $\Upsilon^2=-\theta_v^+$.  Solutions with $k<0$ are incompatible with Eq.~\eqref{thev}. Solutions with $0<k<1$
are excluded by following the chain of reasoning that leads to Eqs.~\eqref{next1}--\eqref{next3} (Appendix \ref{solk}).

For $k=0$ the leading terms in the metric functions in $(t,r)$ coordinates are given  as power series in terms of $x\defeq r-r_\sg$ as
          \be
C= r_\sg-w\sqrt{x}+\frac{1}{3}x\ldots,\qquad h=-\frac{1}{2}\ln{\frac{x}{\xi}}+\frac{4}{3w}\sqrt{x}+\ldots,  \label{k0met}
\ee
 where $w^2\defeq 16 \pi \Upsilon^2 r_\sg^3$ and the higher-order terms depend  on the higher-order terms in the EMT expansion \cite{bmt:18}.
  The function  $\xi(t)$ is   determined by the  choice of the time variable.   The metric functions $C$ and $h$ are obtained as the solutions of Eqs.~\eqref{gtt} and \eqref{grr}, respectively.  Eq.~\eqref{gtr}
  must then hold identically.
Both of its sides of  contain terms that diverge as $1/\sqrt{x}$. Their identification results in the consistency condition
\be
 r'_\sg/\sqrt{\xi}=\pm4\sqrt{\pi}\,\Upsilon\sqrt{ r_\sg}= \pm w/r_\sg.      \label{lumin}
 \ee

A static observer finds that the energy density $\rho=-T^t_{~t}$, the pressure $p=T^r_{~r}$,  and the flux diverge at the apparent horizon.
 On the other hand,   in the reference frame of the infalling observer  on an arbitrary radial trajectory $(T_A(\tau), R_A(\tau),0,0)$ these quantities are
 \be
 \rho_A=p_A=\phi_A=-\frac{\Upsilon^2}{4\dot R^2},             \label{comov}
 \ee
  at the horizon crossing. Additional properties of this metric are discussed in Refs.~\cite{bmmt:18,t:19}

Further relations between the EMT components near the apparent horizon are obtained as follows.
A point on the apparent horizon has the coordinates  $(v,r_+)$ and $(t,r_\sg)$ in the two coordinate systems. Moving from $r_+(v)$ along the line of constant
$v$ by $\delta r$  leads to the point $(t+\delta t, r_\sg+\delta r)$.
 Eqs.~\eqref{intf} and \eqref{lumin} imply
 \be
 \delta t=-\left.\frac{e^{-h}}{f}\right|_{r=r_\sg} \!\!\!\!\! \!\!\!\!\delta r= -\frac{r_\sg\delta{r}}{\sqrt{\xi} w}=\frac{\delta r}{r'_\sg}.
 \ee
  Hence the   EMT components    are related at the first order in $\delta r$ by
\begin{align}
&\pad_r\theta^+_v=-2\Upsilon\Upsilon'/r'_\sg +\alpha, \label{next1}\\
&\pad_r\theta^+_v+\frac{1-w_+}{r_+}\theta^+_{vr}=-2\Upsilon\Upsilon'/r'_\sg+\beta, \label{next2} \\
&\pad_r\theta^+_v+2\frac{1-w_+}{r_+}\theta^+_{vr}=-2\Upsilon\Upsilon'/r'_\sg+\gamma, \label{next3}
\end{align}
where $\pad_r\theta^+_v\defeq\pad_r\theta_v|_{r_+}$, $\alpha(t)\defeq\pad_r \tau_t|_{r_\sg}$, $\beta\defeq\pad_r \tau_t^r|_{r_\sg}$, $\gamma\defeq\pad_r \tau^r|_{r_\sg}$. As a result, the subleading terms satisfy
\be
\alpha+\gamma=2\beta.
\ee


This metric approaches   the pure ingoing Vaidya metric with  decreasing mass, which is the usual near-horizon approximation when the backreaction from  Hawking radiation is taken into account \cite{bardeen:81,bmps:95}.
 The triple limit  $\tau_t,\tau^r,\tau_t^r\to-\Upsilon^2$ is observed in the \textit{ab intio} calculations of the renormalized energy-momentum
tensor on the Schwarzschild background \cite{leviori:16} .

\subsection{Finite density and pressure}

For $k\geqslant 1$ Eqs.~\eqref{fin1} and \eqref{fin2}
do not impose any constraints, and different components of the energy-momentum tensor  can converge to zero at different rates.
However, only  the case $k=1$, where at the leading order in $f$
\be
\tau_t=E(t)f, \qquad \tau^r=P(t)f, \qquad    \tau^r_{t}=\Phi(t) f,   \label{rhop}
\ee
allows for a   solution with $r_\sg'\neq 0$ (see Appendices \ref{bigk} and \ref{firewallB} for details).
These solutions exhibit a finite pressure and a finite density at the apparent horizon, $\rho(t,r_\sg)=E$ and $p(t,r_\sg)=P$, respectively.

Then Eq.~\eqref{gtt} results in the Misner-Sharp mass \be
C=r_\sg(t)+8\pi E r^2_\sg x+\ldots, \qquad  8\pi E r^2_\sg< 1. \label{regas}
\ee
The strict inequality follows from Eq.~\eqref{ric1} below, as $8\pi r_g^2E=1$ is incompatible with $r'_\sg\neq 1$. Consistency of Eqs.~\eqref{gtr} and \eqref{grr} results in
\be
\frac{4\pi (E+P) r^2_\sg}{1-8\pi E r^2_\sg}=-1,
\ee
that ensures the necessary logarithmic divergence  of $h$,
\be
h=-\ln \frac{x}{\xi(t)}+\omega(t)x+\ldots,    \label{hgas}
\ee
for some $\xi(t)>0$ and $\omega(t)$. As a result, Eq.~\eqref{gtr} relates the rate of change of the Schwarzschild radius and the flux as
\be
\qquad r'_\sg=8\pi\Phi\xi r_\sg.
\ee
Requiring the    Ricci scalar to be finite   at $r_\sg$ (Appendix \ref{Ricci}) imposes the constraint
\be
  (1-8\pi E r^2_\sg)\xi=\pm r'_\sg r_\sg,      \label{ric1}
\ee
where  the upper (lower) signs corresponds to the expansion (contraction) of the apparent horizon.

The above conditions imply that a single quantity determines  the two other parameters at the apparent horizon,
\be
P=\frac{-1+4\pi E r^2_\sg}{4\pi r^2_\sg}, \qquad \Phi=\pm\frac{1-8\pi E r^2_\sg}{8\pi r^2_\sg}, \label{out1}
 \ee
where the upper (lower) sign corresponds to accretion (evaporation). The $(t,r)$ block of the EMT is given in Appendix~\ref{firewall1}. The NEC is violated in both cases.
  For example, for $r'_\sg<0$ and the outward pointing null vector $k^\mu$
as $r\to r_\sg$
\be
T_{\mu\nu}k^\mu k ^\nu\approx -\frac{1}{2\pi r_\sg \,x}.
\ee

For  $\Phi>0$  (an accreting  trapped region, $r'_\sg>0$)
 in the reference frame of an infalling observer the energy density
  diverges.  This transient firewall leads to the violation of the QEI, similarly to the $k=0$ case (Appendix \ref{firewall1}).

 The metric of Eq.~\eqref{lfv} describes the $k=1$ evaporating black hole  only if  $w_+\equiv1$.   Compatibility with Eqs.~\eqref{thevr} and  \eqref{ther} results in the relations
\be
  8\pi r_\sg^2(E-\Phi)=1, \qquad E+P-2\Phi=0, \label{ephi}
 \ee
  that are automatically satisfied due to Eq.~\eqref{out1}.

Consider now a time-independent apparent horizon, so the PBH is neither accreting not evaporating, while the solution
 is still time-dependent (such solutions  were considered in the framework of modified gravity, e.g., in \cite{rar:17}).
We treat it as a limiting case of evaporation, $\Phi\leqslant0$. The  condition $r'_\sg=0$ requires
 \be
\frac{4\pi (E+P) r^2_\sg}{1-8\pi E r^2_\sg}=-\lambda,  \label{eqlam}
\ee
$\lambda<1$ to hold. The Ricci scalar   is finite only if  either the density takes the extreme allowed value $E=(8\pi r_\sg^2)^{-1}$, or $\lambda=\half$ (Appendix \ref{Ricci}).  Using Eq.~\eqref{ephi}
 (that still holds up to the end of the dynamical phase),
we obtain
\be
  \Phi=0,\qquad E=-P=1/(8\pi r_\sg^2), \label{statE}  \ee
   in both cases.
   The
NEC is not violated so the solution cannot be realized in finite time $t$.  

 A  static solution    with all metric function being independent of time is possible only if $\tau_t^r\equiv0$. If $h\neq 0$ there is no general requirement $\rho=-p$,  but Eqs.~\eqref{next1}--\eqref{next3}
 imply $E=-P$.

 \subsection{Crossing the apparent horizon}
 Both massless  sufficiently fast  ($4\pi r_\sg^2\Upsilon^2<\dot R^2$) massive test particles cross the apparent horizon of $k=1$ PBH in finite time of a distant observer \cite{t:19}. However,
 it is impossible to fall into a   $k=1$ black hole.

 Consider for simplicity a massless test particle. It is convenient to parameterize the radial ingoing null geodesic $(T_A, R_A)$
 by its radial coordinate,
 $\lambda=-R_A$.   Possibility of the horizon crossing is conveniently monitored by the gap function \cite{kmy:13,nmt:18},
\be
X(\lambda)\defeq R_A-r_\sg\big(T_A(\lambda)\big),
\ee
whose negative rate of change $X_\lambda=dX/d\lambda=-1-r_\sg' dT_A/d\lambda$ indicates that the particle keeps approaching the apparent horizon.

Noting that
 \be
 \frac{dT_A}{d\lambda}= \frac{e^{-h(T_A,R_A)}}{ f(T_A,R_A)}=\frac{r_\sg}{\xi}\frac{1-(\omega- r_\sg^{-1}) X}{1-8\pi E r_\sg^2}+\cO(X^2).
 \ee
Similarly, the rate of change of the coordinate time with repsect to the proper time of an infalling massive test particle is also finite.
   Expanding $X(\lambda)$ in powers of $X$ we find that
\be
X_\lambda=-(\omega- r_\sg^{-1}) X+\cO(X^2).
\ee
If $(\omega- r_\sg^{-1})<0$, then once certain minimal coordinate distance is reached the gap has to increase. If $(\omega- r_\sg^{-1})>0$, then the gap will close exponentially slow,
\be
X\approx X_0 \exp(-\int_{\lambda_0}^\lambda(\omega- r_\sg^{-1})d\lambda),    \label{X38}
\ee
and thus crossing of the apparent horizon ($X=0$) of an evaporating $k=1$ PBH never happens. The same conclusion is obtained by considering a massive test particle and the proper time parametrization.

These results cast doubts on the possibility that $k=1$ black holes can actually form. A thin dust shell, with a flat metric inside and a curved metric outside,  provides the simplest tractable model of the collapse.
The classical Schwarzschild exterior leads to the well-known result  of a finite proper time of the collapse and an infinite collapse time  $t$ according to the clock of a distant observer.
 By using the Vaidya metrics to emulate the effects of evaporation, one  {obtains} results that depend on their choice \cite{bmt:18}.

  By assuming the outgoing Vaidya metric with decreasing mass (which satisfies the NEC and thus cannot lead to
the formation of a PBH in finite  coordinate time $t$),  the apparent horizon is never formed, but the shell either becomes superluminal \cite{cuwy:18}
  or develops a surface pressure at the coordinate distance $x\sim w^2$ from the Schwarzschild radius \cite{bmt:18,nmt:18}.
  On the other hand,   the  {ingoing Vaidya} metric of Eq.~\eqref{lfv} leads   to the horizon formation in finite time according to both clocks \cite{bmt:18}.    However, if the exterior is modelled  by Eq.~\eqref{sgenm}
  with $k=1$ metric functions \big(Eqs.~\eqref{regas} and \eqref{hgas}\big),  Eq.~\eqref{X38} indicates that the shell's collapse will never be complete,
  even if the exterior metric violates the NEC.

%

                              \section{Implications for models of  regular black holes}

Whether their motivation is to construct a geodetically complete spacetime, to resolve the information loss paradox, or to illustrate the effects of quantum gravity,
 models of regular black holes (RBHs) envisage a trapped region with a singularity-free core (see e.g.,~\cite{bar:68,fv:81,hay:06,f:14,bm:19,regular} and the reviews \cite{coy:15, rev:08,math:05}).
 Considerations of a geometric nature \cite{pandora-4}, as well as constrains from the effective field theory of  quantum gravity \cite{fmm:89,dlprs:15} restrict these  models.
 Here we explore the further constraints that are imposed by the results of Sec.~II.

  \begin{figure}[htbp]
\includegraphics[width=0.33\textwidth]{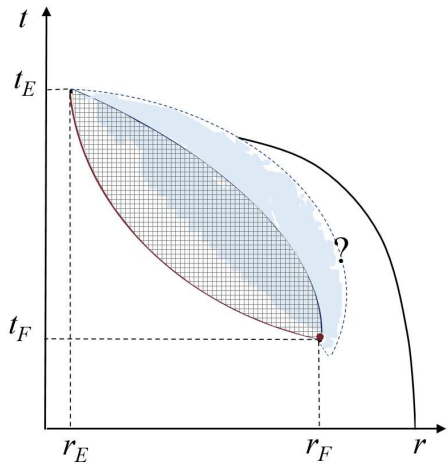}
\caption{{ Schematic depiction of the evolution of a RBH from the point of view of a distant observer.  The dark blue line represents the apparent horizon and the double dark red line represents the inner horizon.
The trapped region is cross-hatched. The NEC-violating region (blue spread, dashed boundary) appears prior to the formation of the first marginally trapped surface $(t_F,r_F)$
  and covers part of the trapped region. Its outer boundary is not constrained by our considerations. The thin black line traces the surface of the collapsing body  up to the NEC-violating region.
   The trapped {region} evaporates at some $(t_E,{r_E})$ where the two hypersurfaces cross again. }}
\label{schema1}
\end{figure}

 Many of the proposed static models \cite{bar:68,hay:06,regular} assume finite density and pressure at the horizon and thus belong to the class $k=1$. However, such solutions cannot be realized without  a
   breakdown of the semiclassical physics.  Leaving aside the doubts about viability of the dynamical $k=1$ solutions, the static situation ($E=-P$, $\Phi=0$) still cannot
 arise at finite $t$, as in this case the NEC is satisfied and the apparent horizon is hidden from the distant observer by the event horizon.
 Appearance of the feature that the model is built to prevent indicates its breakdown.

Even the asymptotic case cannot be realized without some radical departures from the semiclassical physics.  The zero flux limit can be produced only if the scenario of Eq.~\eqref{statE} is realized.
However,  in the limit $8\pi r_\sg^2E\to 1$ the formerly regular terms in the curvature scalars diverge (Appendix \ref{Ricci}).

   The leading behaviour of the function
 $h$ of the  $k=0$ solutions matches the regular static scenario with $k=1$, $\lambda=\half$. Nevertheless, the latter  is not a suitably defined limit of the former. First,
 to produce this effect some mechanism should freeze the apparent horizon and thus push $\Upsilon$ in  $\tau_t=-\Upsilon^2+\alpha x+\ldots$, etc., to zero. This is exactly opposite of the expected
  semiclassical behavior \cite{fn:book,rev-1,bmps:95}.
Moreover,  after the freezing,   to avoid a  discontinuous change in the black hole (Misner-Sharp) mass,
  the linear terms in Eqs.~\eqref{k0met} and \eqref{regas} should match.  This leads to a contradiction
  \be
 E=\frac{1}{24\pi r_\sg^2}\neq \frac{1}{8\pi r_\sg^2}=-P,
  \ee
where the first value of $E$ is obtained by matching with the linear part of Eq.~\eqref{k0met} and the second value results from Eq.~\eqref{statE}.

A dynamical model of Hayward and Frolov \cite{hay:06,f:14} uses $(v,r)$ coordinates    and the minimal modification of the Vaidya metric by setting
\be
 C_+(v)=\frac{2m(v)r^3}{r^3+2m(v)b^2}, \quad h_+=0,  \label{hf-m}
\ee
for some $b>0$ and  $m(v)$.
      When $m\gg b$ the approximate locations of the apparent horizon and the inner horizon are given by
 \be
 r_\sg\approx 2m-\frac{b^2}{2m}, \qquad r_\mathrm{in}\approx \frac{5b}{4}-\frac{3b^2}{32m},    \label{horizons}
 \ee
 respectively, and the non-zero components of the energy-momentum tensor at the apparent horizon are
\be
\Theta_{vv}\approx \frac{m'(v)}{16 \pi m^2(v)}, \qquad \Theta_{vr}\approx -\frac{3b^2}{128 m^4}.
\ee
This model belongs to $k=0$ class. It is consistent with formation of the apparent horizon  at a finite time of a distant observer.

However, it  is a consistent description of only the evaporation part of the RBH evolution and cannot describe its formation. Leaving aside the issue of a transient firewall that accompanies accretion,  for $m'(v)>0$ the NEC is not violated in this model. Thus the apparent horizon, if exists, is hidden behind the even horizon that was purportedly eliminated.
   In fact, no model that uses $(v,r)$ coordinates and has a regular function $h_+(v,r)$ can describe growth of a PBS, as in this case $\tau_t^r \to+\Upsilon^2$
\be
\frac{\pad_r h_+}{r}=4\pi\Theta_{rr}^+\to\frac{16\pi}{f^2}\Upsilon^2,
\ee
that ensures divergence of at least of $\pad_r h_+$.

Since the energy density and pressure are negative in the vicinity of the apparent horizon and positive in the vicinity of the inner horizon \cite{frolov-do,t:19} there should be density and pressure
 jumps at the intersection   of the two horizons, making problematic the blanket requirement of continuity of density and pressure.
  If we accept that violations of the QEI is a sufficient reason to discount the growth of trapped region, the horizon structure of a regular black hole is schematically shown on Fir.~1.

In this case the model with the metric functions~\eqref{hf-m} cannot describe  the first stages of the evolution of the trapped region, even if $C'_+<0$. For a RBH of Fig.~1
both the apparent horizon and the inner
 horizon develop from      a single trapped surface that appears at some $t_F$ and meet again at $t_E$, possibly forming a remnant. The Misner-Sharp mass  of Eq.~\eqref{hf-m} allows a latter possibility
 (at $m(v_E)=3\sqrt{3}b/4$),
 but not the former one, as Eq.~\eqref{horizons} indicates.

\section{Discussion}
      We have seen that in the vicinity of the apparent horizon a singular nature of Schwarzschild coordinates serves a useful purpose.   Scaling of the suitably selected functions of the EMT
  components with the powers $k$ of $f=(1-C(t,r)/r)$ allows to classify solutions of the Einstein equations. Only two types of solutions with $k=0,1$ are possible. Both violate the NEC  and result in a firewall
  at the expanding apparent horizon. Only
  $k=0$ solutions allow to a collapsing thin shell to form a black hole or for a test particle to cross the apparent horizon. These failures cast doubts on the physical relevance of the $k=1$ solutions.

 Analysis of the inner regions of RBHs leads to the arguments indicating the need for physics beyond standard model to support such objects \cite{bm:19,bm:19b}.
  Our analysis of the near-horizon regions indicates that $k=0$ models
 of evaporating RBHs are  as exotic as  any UCO with or without an apparent horizon. On the other hand, complete regularity
  (finite values of density and pressure for both static and   infalling observers) of $k=1$ PBH may be impossible to realize without significant modification of the semiclassical gravity.

   \acknowledgments  Useful discussions with Valentina Baccetti, Robert Mann and Sebastian Murk are gratefully acknowledged.

\appendix

\section{Solutions with $k<1$}

 \subsection{Behaviour of $\mathbf T^\theta_{\,\theta}$}\label{apreg}

 The regularity conditions   Eqs.~\eqref{fin1} and \eqref{fin2}
\begin{align}
\mathrm{T}&=-\frac{\tau_t}{f}+\frac{\tau^r}{f}+2  T^\theta_{\,\theta}, \\
 \mathfrak{T}&=\left(\frac{\tau_t}{f}\right)^2+  \left(\frac{\tau^r}{f}\right)^2-2\left(\frac{\tau_t^r}{f}\right)^2  +2\big(T^\theta_{\,\theta}\big)^2,
  \end{align}
constrains the leading term in $T^\theta_{\,\theta}\equiv T^\phi_{\,\phi} $ when $r\to r_\sg$ if the three other components of the EMT scale as $f^k$, $k<1$. Set
 \begin{align}
 &\Xi_1\defeq\lim_{r\to r_\sg}\tau_t/f^k, \qquad &\Xi_2\defeq\lim_{r\to r_\sg}\tau^r/f^k, \\
 &  \Xi_3\defeq\lim_{r\to r_\sg}T^\theta_{\,\theta}/f^{k-1}, \qquad  &\Xi_4\defeq\lim_{r\to r_\sg}\tau_t^r/f^k,
 \end{align}
and focus on the leading terms. The two conditions become
  \be
-\Xi_1+\Xi_2+2\, \Xi_3=0, \qquad  \Xi_1^2+\Xi_2^2+2\,\Xi_3^2-2\,\Xi_4^2=0,
\ee
Taking $\Xi_1$ and $\Xi_2$ as the independent variables we find
\be
\Xi_3=\half(\Xi_1-\Xi_2), \qquad \Xi_4=\pm\frac{1}{2}\sqrt{3\, \Xi_1^2+3\, \Xi_2^2-2\,\Xi_1\Xi_2}.
\ee
 The Einstein equation \eqref{gtt} does not change; the leading terms of the Misner-Sharp mass are
\be
C=r_\sg-w_1 x^{1/(2-k)},
\ee
where $w_1=\big(-8(2-k)\pi r_\sg^{3-k}\Xi_1\big)^{1/(2-k)}$. The limiting form of Eq.~\eqref{grr} now becomes
    \be
    \pad_rh= 4\pi(\Xi_1+\Xi_2)r_\sg^{3-k}\frac{w_1^{k-2}}{x}=-\frac{\Xi_1+\Xi_2}{2(2-k)\Xi_1 x},
    \ee
   that results in the leading term
    \be
h=-\frac{\Xi_1+\Xi_2}{2(2-k)\Xi_1 }\ln\frac{x}{\xi}.
\ee

Hence consistency of Eq.~\eqref{gtr} imposes \be
 -\frac{\Xi_1+\Xi_2}{2(2-k)\Xi_1 }=-\frac{1}{2-k}, \ee
resulting in  $\Xi_1=\Xi_2$ and
\be
\Xi_3=0, \qquad \Xi_4=\pm \Xi_1.
\ee

 \subsection{Regular solutions in $(v,r)$ coordinates.}  \label{rvr}
 Existence of the apparent horizon constrains the Misner-Sharp function to have the form
 \be
 C_+=r_+(v)+w_1^+(v)x^{1-\alpha_1}+w_2^+(v)x^{1-\alpha_1+\alpha_2}+\ldots,
 \ee
where $x=r-r_+$, and   $\alpha_1<1$, $\alpha_2>0$,
while we keep the function $h$ unconstrained,
 \be
 h_+=h_+(v)\ln x/\xi(v)+h_1^+(v)x^\beta_1+  h_2^+(v)x^{\beta_1+\beta_2}+\ldots.
 \ee
Explicit evaluation of the Ricci scalar $R$ for the  metric \eqref{lfv} with  the above functions results in a number of the divergent terms, that as $x\to 0$ behave as its various powers.
 The curvature scalar is finite if the coefficients of all such divergent powers cancel. However, it is possible only if $h_+=0$, as well as all the coefficients of all fractional powers that are less than two.

 \subsection{Leading terms of the solutions, $k<1$} \label{solk}
For the EMT components of Eq.~\eqref{taus} with $k<1$ (and thus finite $T^\theta_{\,\theta}\equiv T^\phi_{\,\phi}$) the Einstein equations with divergent terms become
\begin{align}
&\pad_r C \approx-8\pi r_\sg^2 {\Upsilon^2} {f^{k-1}}, \label{gttA}\\
& \pad_t C \approx\pm 8\pi r_\sg^2 e^h\Upsilon^2 f^k, \label{gtrA}\\
& \pad_r h\approx-8\pi r_\sg \Upsilon^2 f^{k-2}. \label{grrA}
\end{align}
The sign choice follows from the observation that the equations have no real solutions if $\tau_t$  and $\tau^r$ are $+\Upsilon^2 f^k$. The leading terms of the metric functions are then
  \be
  C=r_\sg(t)-w_1 x^{1/(2-k)}, \qquad w_1^{2-k}=8(2-k)\pi r_\sg^{3-k}\Upsilon^2,      \label{Ck}
   \ee
  and
  \be
  h=-\frac{1}{2-k}\ln\frac{x}{\xi}.     \label{hk}
   \ee
Eq.~\eqref{gtrA} results in the constraint
  \be
\pm r'_\sg=\frac{w_1 \xi^{1/(2-k)}}{r_\sg}.
  \ee

It was shown in Section \ref{divp} that solutions with $k<0$ are incompatible with Eq.~\eqref{thev}.  By following the chain of reasoning that established
Eqs.~\eqref{next1} -- \eqref{next3}  we show that solutions with $k>0$ are also inadmissible.   For $k<1$ moving from $\big(v,r_+(v)\big)$ along the line of constant
$v$   leads to the point $(t+\delta t, r_\sg+\delta r)$, where
 Eqs.~\eqref{Ck} and \eqref{hk} imply
 \be
 \delta t=-\left.\frac{e^{-h}}{f}\right|_{r=r_\sg} \!\!\!\!\! \!\!\!\!\delta r=\frac{\delta r}{r'_\sg}.
 \ee
For $k\neq 1$ the analog of Eq.~\eqref{next1} is a contradictory expression \be \pad_r\theta^+_v=-\frac{k\Upsilon^2}{f^{1-k}}+\ldots\rightarrow -\infty, \ee showing that solutions with $0<k<1$ are to be excluded.

\subsection{Leading terms of the solutions, $k\geqslant 1$} \label{bigk}
For the EMT components of Eq.~\eqref{taus} with $k\geqslant 1$ the Einstein equations with divergent terms become
\begin{align}
&\pad_r C \approx 8\pi r_\sg^2 {E(t)} {f^{k-1}}, \label{gttA1}\\
& \pad_t C \approx 8\pi r_\sg^2 e^h \Phi(t) f^{k_\Phi}, \label{gtrA1}\\
& \pad_r h\approx 4\pi r_\sg\big( E(t) f^{k-2}+P(t) f^{k_P-2}\big), \label{grrA1}
\end{align}
for some functions $E(t)$, $P(t)$ and $\Phi(t)$ and powers $k,k_\Phi,k_P\geqslant1$.
 The leading terms of the Misner-Sharp mass are then
  \be
  C=r_\sg(t)+8\pi E r_\sg^2 x^k. \ee
   For $k>1$ the constraints of Sec II.B do not apply.     In this case
   \be
   f=\frac{x}{r_\sg}+\ldots.
   \ee

Solutions with variable $r_\sg(t)$ impose via Eq.~\eqref{gtrA} that $e^h\propto x^{-k_\Phi}$, i. e. the logarithmic divergence of the function $h$. For $k>1$ it can be realized only if $k_P=1$. Then
\be
  h=4\pi P r_\sg^2 \ln \frac{x}{\xi}, \qquad   4\pi P r_\sg^2=-k_\Phi.
   \ee
Further properties of these solutions are discovered by using the relations between the EMT components \eqref{thev}--\eqref{ther}.
  Eq.~\eqref{thev} is satisfied if $w_+=1$. Eq.~\eqref{thevr} then implies $k_\Phi=1$ and $\Phi=-1/(8\pi r_\sg^2)$.    Eq.~\eqref{ther} is satisfied if $k\geqslant 2$. We see that these solutions are rather
  peculiar:   energy density vanishes at the apparent horizon and the pressure and the flux are determined by the Schwarzschild radius. Moreover, the firewall  is present even
  if $\Phi<0$ (Appendix  \ref{firewallB}).

   For $k=1$ and $k_\Phi>1$  the equality $8\pi E r_\sg^2=1$ is still impossible.  Assuming that it is true we find that the Misner-Sharp mass  in the vicinity of $r_\sg$ is
    \be
C=r_\sg+x-b^2 x^2+\ldots, \qquad f=b^2 x^2/r_\sg+\ldots,
\ee
for some $b(t)$.     Eq.~\eqref{gtr} becomes in the leading order
\be
\frac{2b^2 r'_\sg \,x}{r_\sg} =8 \pi \Phi \left(\frac{b^2 x^2}{r_\sg}\right)^{k_\Phi} \!\!\!e^h,  \ee
requiring
\be
h=-(2k_\Phi-1)\ln\frac{x}{\xi}+\ldots,
\ee
where the higher-order terms are omitted, for time-dependent Schwarzschild radius.
Eq.~\eqref{grr} then results in the leading order  relation
\be
\frac{\pad_ x h}{r_\sg}=  \frac{r_\sg^2}{b^4 x^4}\left(E \frac{b^2 x^2}{r_\sg}+P\left(\frac{b^2 x^2}{r_\sg}\right)^{k_P}\right),
\ee
resulting in $1/x$ divergence of the function $h$.

Evaluation of the Ricci scalar with these metric functions results in the divergent expression unless    $k_P=k_\Phi=1$.     It is given in Appendix \ref{Ricci}.

         \section{Some properties of the solutions}
 \subsection{Firewall at the apparent  horizon, $k<1$} \label{firewall}
                     For a radially infalling massive particle the four-velocity components are related by
                     \be
                     \dot T_A=\frac{\sqrt{F+\dot R_A^2}}{e^HF}\approx \frac{|\dot R|}{e^HF}+  \frac{1}{2|\dot R_A|e^H},
                     \ee
where $H=h(T_A,R_A)$ and $F=f(T_A,R_A)$. For $k<$ this means that the four-velocity of an infalling observer at the leading order is given by
 \be
u_A^\mu=|\dot R_A|\left( \frac{r_\sg}{ w_1 \xi^{1/(2-k)} },-1,0,0\right),
\ee
while the leading terms in  the $(t,r)$ block of the EMT are
\be
T_{ab}=-\frac{\Upsilon^2}{x}\frac{ w_1^{k-2}}{r_\sg^{k-2}}\begin{pmatrix}
w_1^2\xi^{2/(2-k)}/r_\sg^2 &\pm w_1 \xi^{1/(2-k)}/r_\sg \vspace{1mm}\\
\pm w_1 \xi^{1/(2-k)}/r_\sg  & 1 \end{pmatrix},
\ee
where the upper (lower) sign corresponds to $r'_\sg<0$ ($r'_\sg>0$), respectively. The energy density in the frame of the particle is $\rho_A=T_{\mu\nu}u_A^\mu u_A^\nu$.

For $r'_\sg<0$ the divergent terms in the energy density  cancel out.
 However, for the expanding apparent horizon the energy density is negative and divergent,
\be
\rho_A\approx -\frac{4 \dot R^2 \Upsilon^2 r_\sg^{2-k}}{w_1^{2-k}X},
\ee
where $X=R_A-r_\sg$.

      \subsection{Firewall at the apparent  horizon, $k=1$.} \label{firewall1}

The leading terms of the  metric functions are
\be
C=r_g+8\pi E r_\sg^2 x, \qquad h= -\ln x/\xi.
\ee
The four-velocity of an infalling observer at the leading order is then  given by
 \be
u_A^\mu=|\dot R_A|\left( \frac{r_\sg}{\xi(1-8\pi E r_\sg^2) },-1,0,0\right),
\ee
  and the leading terms in  the $(t,r)$ block of the EMT are
\be
T_{ab}=\frac{1}{r_\sg\, x} \!\!\begin{pmatrix}
8\pi E(1-8\pi Er_\sg^2)\xi^2 & \hspace{-10mm}r_\sg' \vspace{1mm}\\
r_\sg' & \hspace{-10mm} -{(2-8\pi Er_\sg^2)}/(1-8\pi Er_\sg^2)\vspace{1mm} \end{pmatrix}.
\ee
 In the case of expansion  the function $\xi$ satisfies
 \be
 (1-8\pi Er_\sg^2)\xi=+r'_\sg r_\sg, \ee
 and the resulting energy density diverges as
\be
\rho_A \approx -\frac{4 \dot R^2}{r_\sg X}.
\ee

 For spacetimes of small curvature explicit expressions that bound time-averaged energy density for a geodesic observer
  were derived in Ref.~\cite{eleni}.
   For any Hadamard state $\omega$ and a sampling
function $\mathfrak{f}(\tau)$ of compact support,
 negativity of the expectation  value of the energy density  $\rho_A$ as seen by a geodesic observer  on a trajectory $\gamma(\tau)$ is bounded by
\be
 \int_\gamma\! \mathfrak{f}^2(\tau)\rho d\tau \geqslant - B(R,\mathfrak{f},\gamma), \label{qei}
\ee
where $B>0$ is a bounded function that depends on the trajectory, the Ricci scalar and the sampling function \cite{eleni}.

Consider a growing apparent horizon, $r_\sg'>0$.
 For a macroscopic black hole the curvature at the apparent
 horizon is low and its radius does not appreciably change while the observer (a massive test particle) moves in its vicinity. Then $\dot X \approx \dot R$, and for a given geodesic trajectory
  we can choose  $\mathfrak{f}\approx 1$ at
  the horizon crossing and $\mathfrak{f}\to 0$  within the NEC-violating domain. As the trajectory
 passes through  $X_0+r_\sg\to r_\sg$ the lhs of Eq.~\eqref{qei} behaves as
 \be
 \int_\gamma\! \mathfrak{f}^2\rho_\mA d\tau\approx -\int_\gamma\frac{4\dot R^2d\tau}{r_\sg\, X}\approx \int_\gamma\frac{4|\dot R|dX} {r_\sg\, X}\propto \log X_0\to-\infty,
 \ee
where we used $\dot R\sim\mathrm{const}$. The rhs  of Eq.~\eqref{qei} remains finite, and thus the QEI is violated.

\subsection{Firewall at the apparent  horizon, $k>1$.} \label{firewallB}
Using the results of Appendix \ref{bigk} for $k>1$, $k_P=k_\Phi=1$ we find that
the four-velocity of an infalling observer at the leading order is then  given by
 \be
u_A^\mu=|\dot R_A|\left( \frac{r_\sg}{\xi },-1,0,0\right),
\ee
  and the leading terms in  the $(t,r)$ block of the EMT are
\be
T_{ab}=\begin{pmatrix}
 E & -(8\pi r_\sg^2)^{-1} \vspace{1mm}\\
-(8\pi r_\sg^2)^{-1} & -(4\pi r_\sg x)^{-1} \vspace{1mm} \end{pmatrix},
\ee
where the minimal allowed power $k=2$ was used in $\tau_t=E f^k$.

As a results the energy density in the infalling frame \be
\rho_A \approx -\frac{1}{4 \pi r_\sg X},
\ee
diverges even for an evaporating $k\geqslant 2$   black hole, and the violation of the QEI is established analogously to   Appendix \ref{firewall}.

\subsection{The Ricci scalar, $k=1$.} \label{Ricci}

For a dynamical solution in the case $k=1$ with the metric functions given by Eqs.~\eqref{regas} and \eqref{hgas} expansion of the Ricci scalar near the apparent horizon gives
\be
R=-\frac{(1-8\pi Er_\sg^2)^2\xi^2-r_\sg^2r'_\sg{}\!^2}{r_\sg(1-8\pi Er_\sg^2)\xi^2 x}+\cO\big(x^0\big). \label{b12}
\ee
It is finite if and only if Eq.~\eqref{ric1} is satisfied.

The Ricci scalar diverges if the evaporating black hole freezes ($r_\sg\to 0$), as the regular (as a function of $x$)  part of $R$ contains a clearly divergent term
\be
R_0=\frac{1}{r_\sg^2 r_\sg'}.
\ee

If the metric function $h$ has a proportionality coefficient that is different from one (if  $k_\Phi>1$ while $k_P\geqslant 1$ and $k=1$),
\be
h=-\lambda\ln \frac{x}{\xi}+\ldots, \label{hlam}
\ee
Eq.~\eqref{gtr} implies that $\lambda=k_\Phi$, and the Ricci scalar  contains a  potentially divergent term \be
R_{-1}=\frac{\lambda(2\lambda-1)(1-8\pi E r_\sg^2)}{r_\sg \,x},
\ee
that will be zero only if $\lambda=0,\half$ or $E=(8\pi r_\sg^2)^{-1}$.

The two former options (with $8\pi Er_\sg^2<1$) contradict the assumption $\lambda=k_\phi>1$.
The third option --- the identity $8\pi Er_\sg^2=1$ --- is impossible to satisfy, as demonstrated in Appendix \ref{bigk}.

\end{document}